\documentclass{osa-article}
\journal{oe}

\usepackage{graphicx}
\usepackage[separate-uncertainty=true]{siunitx}
\DeclareSIUnit{\belmilliwatt}{Bm}
\DeclareSIUnit{\dBm}{\deci\belmilliwatt}
\usepackage{physics}

\begin{document}

\title{Reconfigurable frequency coding of triggered single photons in the telecom C--band}

\author{Samuel Gyger,\authormark{1, 4, 5} 
Katharina D. Zeuner,\authormark{1, 4}
Klaus D. J\"{o}ns,\authormark{1} 
Ali W. Elshaari,\authormark{1} 
Matthias Paul,\authormark{1} 
Sergei Popov,\authormark{1} 
Carl Reuterski\"{o}ld Hedlund,\authormark{3} 
Mattias Hammar,\authormark{3} 
Oskars Ozolins,\authormark{1, 2, 6} 
and Val Zwiller\authormark{1} }

\address{\authormark{1}KTH Royal Institute of Technology, Department of Applied Physics, Albanova University Centre, Roslagstullsbacken 21, 106 91 Stockholm, Sweden\\
\authormark{2}NETLAB, RISE AB, Isafjordsgatan 22, 164 40 Kista, Sweden\\
\authormark{3}KTH Royal Institute of Technology, Department of Electronics, Electrum 229, 164 40 Kista, Sweden\\
\authormark{4}Equal Contribution\\
\authormark{5}\email{gyger@kth.se}
\authormark{6}\email{oskars.ozolins@ri.se}}
\medskip
\begin{abstract}
In this work, we demonstrate reconfigurable frequency manipulation of quantum states of light in the telecom C--band. Triggered single photons are encoded in a superposition state of three channels using sidebands up to  \SI{53}{\giga\hertz} created by an off--the--shelf phase modulator. The single photons are emitted by an InAs/GaAs quantum dot grown by metal-organic vapor-phase epitaxy within the transparency window of the backbone fiber optical network. A cross-correlation measurement of the sidebands demonstrates the preservation of the single photon nature; an important prerequisite for future quantum technology applications using the existing telecommunication fiber network.
\end{abstract}

\section{Introduction}
Quantum information science is an interdisciplinary field wherein quantum mechanics sets rules for logical bits (qubits), gates and interconnects \cite{DiVincenzo1995}. Transferring quantum information between stationary nodes is one of the requirements to scale the field beyond isolated locations\cite{Kimble2008}. 

Qubit teleportation protocols \cite{Bouwmeester1997, Boschi1998, Pirandola2015, Reindl2018} or direct state transfer methods\cite{Sherson2006, Gao2013} between two nodes are commonly built based on single photons, so called "flying qubits". Photons at a wavelength of $\SI{1.55}{\micro\meter}$\cite{Miya1979} (C--band) can travel over large distances in optical fibers and are ideal candidates to distribute entanglement or exchange quantum information between distant network nodes, known as "quantum internet"\cite{Kimble2008, Wehner2018}. The advantage of pre--existing infrastructure for optical classical communication networks\cite{Agrawal2010} leads to faster and cheaper adoption of quantum communication.

Semiconductor quantum dots (QDs) are promising sources for flying qubits at \SI{1.55}{\mu\meter}\cite{Chithrani2004, Paul2017, Muller2018}, due to the deterministic generation of single photons\cite{He2013} and polarization entangled photon pairs\cite{Akopian2006, Olbrich2017}. The quantum dot growth via metal--organic vapor--phase epitaxy enables industrial large--scale fabrication in future photonic quantum technology applications. By combining this industry grade growth technique, used for example in LED manufacturing, with nanofabrication methods, QDs establish themselves as scalable, integratable\cite{Davanco2017} and tunable sources for the C--band\cite{Zeuner2018}. As compared to the standard InP--based materials for this wavelength range, the choice of InAs/GaAs quantum dots has several advantages regarding the optical properties of the quantum dots and due to the possibility to grow lattice-matched DBRs with high refractive index contrast materials, the possibilities for cavity--enhanced emission enabling ultra-high repetition rates\cite{Paul2017, Senellart2017}.

Building up multi--node quantum networks requires multiple sources which need to be tuned in resonance to one another within their natural linewidth or wavelength-multiplexed in the same fiber for enhanced quantum communication bandwidth and multiple access options. Thus, dynamically manipulating single photon properties such as frequency\cite{Lo2017}, line shape \cite{Belthangady2010} or temporal shape \cite{Kolchin2008,Specht2009} is a key requirement for quantum information science.  Although, attenuated coherent carriers with sidebands were proposed and used for Quantum Key Distribution \cite{Merolla2002}, moving to flying qubits with high purity number states enables new applications, e.g. phase--modulated single photons are suggested for frequency--based linear optical quantum computing \cite{Lukens2017} and frequency-based higher order entangled states\cite{Kues2017}. Constructing a flying qubit Hilbert space in frequency domain is bound to boost quantum data communication bandwidth, following the steps of classical optical communication where generally a combination of time-division coding and frequency-division coding is used for high data-rate applications. The possible on-demand generation\cite{He2013} of single photons with InAs/GaAs QD coupled with the frequency coding\cite{Paudel2018} we demonstrate here, using off-the-shelf components, makes the platform ideal for long distance and high bandwidth quantum communication.

Paudel et al.\cite{Paudel2018} recently demonstrated the preservation of indistinguishability of near-infrared photons emitted by a quantum dot through phase modulation. We extend this result and move to a triggered QD system with an emission in the standard telecom \SI{1.55}{\mu\meter} band. We experimentally demonstrate triggered generation, manipulation and transmission of single photons in the telecom C--band. We create a superposition state spread over three adjacent channels ($\SI{26.5}{\giga\hertz}$ separation), which we then transfer over data--center distance ($\SI{1.6}{\kilo\meter}$). The transmission link consists of an InAs/GaAs quantum dot source, an off-the-shelf phase modulator and telecom single mode fibers (SMF--28). We show that the sideband photons generated with the phase modulator follow the properties of the unmodulated photons emitted by a single QD. Our work demonstrates the integration of novel single photon sources with mature telecom technology, allowing large--scale quantum information processing and networking in the \SI{1.55}{\mu\meter} band.

\begin{figure}[htbp]
    \includegraphics[width=140mm]{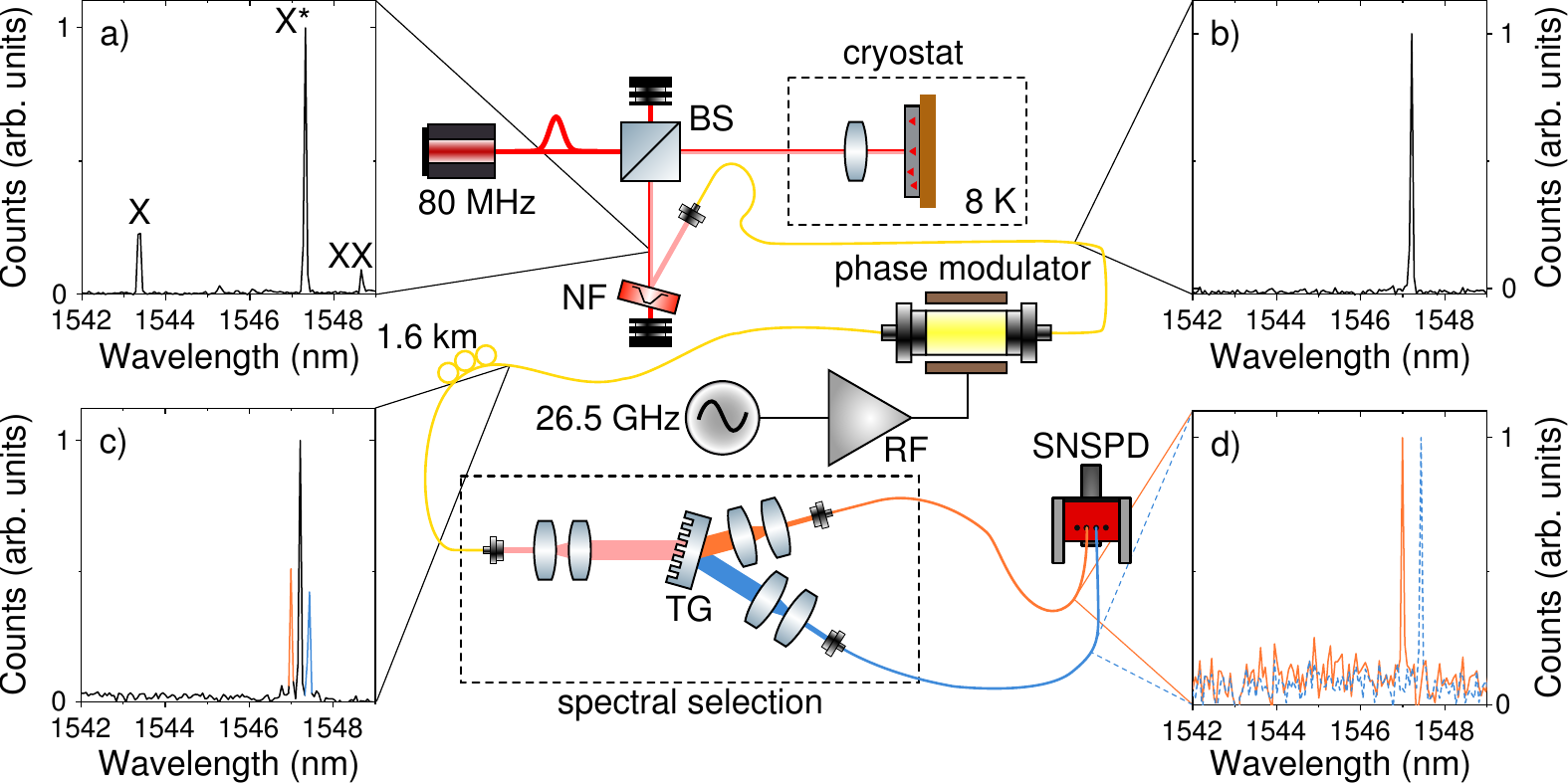}
    \caption{\label{fig:figure1}The setup with spectra at different positions in the experiment. Single photons of a MOVPE--grown InAs/GaAs QD in a closed--cycle cryostat (a) are fiber coupled through a confocal microscope. (b) A notch--filter (NF) is used to reflect one emission line of the quantum dot. (c) The reflected single photons are sent to a phase modulator driven with $\SI{26.5}{\giga\hertz}$, (d) spectrally filtered using a transmission spectrometer and connected to superconducting nanowire single photon detectors (SNSPD).}
\end{figure}
\section{Experiment}
Our experimental setup is described in Fig.~\ref{fig:figure1}. The source used in the experiment is based on MOVPE grown InAs quantum dots on a GaAs substrate. A distributed Bragg reflector consisting of 20 pairs of AlAs/GaAs is used for improved extraction efficiency and a metamorphic buffer layer made from GaAs with increasing In content is providing stress--release for the InAs quantum dots and, thus, allowing emission in the telecom C-band. A more detailed description of the sample can be found in \cite{Zeuner2018}. During the experiment the sample is placed in a close--cycle cryostat with a base temperature of \SI{8}{\kelvin}, and a microscope objective (\SI{0.85}{NA}) as part of our confocal microscopy setup.
The quantum dot is excited at \SI{1470}{\nano\meter} with \SI{2}{\pico\second} pulses and a repetition rate of \SI{80}{\mega\hertz}. We expect our combined extraction efficiency and setup collection efficiency to be on the order of $10^{-4}$.
Based on spectral measurements, only a specific quantum dot emission line is selectively reflected from a tunable notch filter with a spectral bandwidth of \SI{0.7}{\nano\meter} and a rejection of \SI{40}{\decibel}. 
The filtered single photons are then coupled in an optical fiber connected to the modulation setup. We use a standard LN phase modulator (Sumitomo T.PM1.5--40) with a $V_\pi=\SI{5}{V}$ to create the sideband photons. The modulator is driven at \SI{5}{Vpp} to create sidebands at a spectral distance of \SI{26.5}{\giga\hertz}.
The modulated photons are sent through a \SI{1.6}{\kilo\meter} long SMF--28 fiber (\SI{0.2}{\decibel\per\kilo\meter} attenuation) before the generated sidebands are separated from the carrier using a home--built transmission spectrometer\cite{Schweickert2018}. 
After passing through the transmission grating (TG) the two sidebands (spectrally separated by \SI{0.4}{\nano\meter}) are coupled into two optical fibers that are connected to superconducting nanowire single photon detectors (\SI{15}{\percent} and \SI{25}{\percent} efficiency with \num{30} dark counts per second). 
To record the correlations between the single photon sidebands a correlator (\SI{50}{\pico\second} time jitter between channels) is used. Additionally, a spectrometer (\SI{750}{\milli\meter} focal length, 830\,l/mm grating) and an InGaAs array are used for the spectral analysis.
\begin{figure}[htbp]
    \includegraphics[width=138mm]{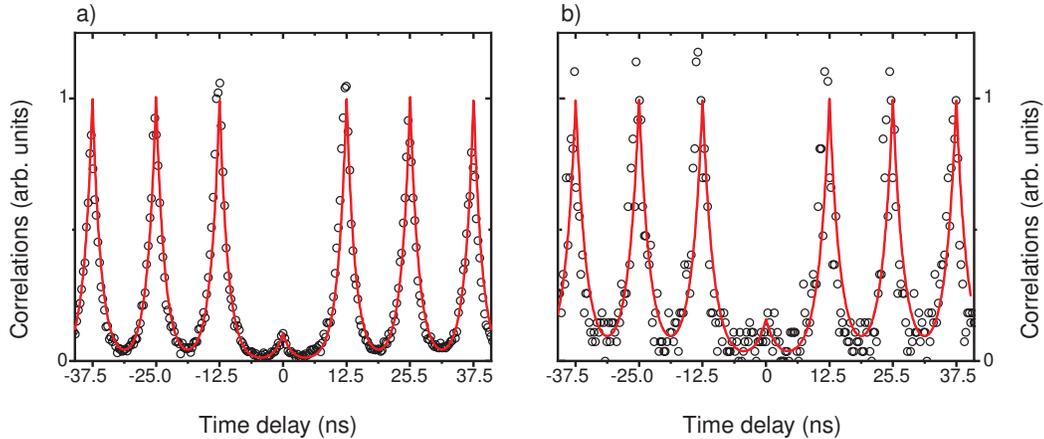}
    \caption{\label{fig:figure2}(a) Auto--correlation measurement on the photons emitted into the charged exciton line X* at \SI{1547.21}{\nano\meter} filtered by the notch--filter yielding a multi--photon probability of $g^{(2)}(0)=\num{0.11 \pm 0.03}$.  (b) Cross--correlation measurement between the two sidebands at \SI{1546.99}{\nano\meter} and \SI{1547.42}{\nano\meter} yielding a multi--photon probability of $g^{(2)}(0)=\num{0.16 \pm 0.06}$. The open circles represent the measurement data, the solid lines correspond to a fit to the data. The presented data is not background subtracted and normalized to the peak height provided by the fit.}
\end{figure}
\section{Results}
In a QD, the emission is related to the particles involved in the recombination process. A systematic study, including power and polarization dependent measurements \cite{Zeuner2018}, was performed to identify three optical transitions X, XX, and X* in the spectra. After excitation, the s--shell of our source can be filled with two electrons in the conduction band and two holes in the valence band. This can result in the emission of a biexciton photon via the recombination of one electron--hole pair, labeled XX. The remaining electron and hole pair in the QD recombines to emit an exciton photon, labeled X. Additionally, the QD can trap a single electric charge in the presence of an electron--hole pair, the three-particle complex produces a trion, labeled X*.  A spectrum of our source is shown in Fig.~\ref{fig:figure1}(a). All spectra are normalized to their peak value. We select the trion (X*) line emitted by the quantum dot by reflecting only this wavelength (\SI{1547.21}{\nano\meter}) on the tunable notch filter. The filtered spectrum is depicted in Fig.~\ref{fig:figure1}(b). We would like to note, that the common insertion loss of commercial telecom equipment on the order of at least 3dB per element, which can be easily compensated in classical communication, is a significant challenge for experiments with single photons. 

The multi-photon emission probability of our source is measured by performing a Hanbury Brown and Twiss measurement. To this end, we send the filtered X* photons onto a beam splitter and measure the difference in arrival time between the beam splitter outputs with two superconducting nanowire single photon detectors. This yields the histogram shown in Fig.~\ref{fig:figure2}(a).

The absence of the peak at zero time delay demonstrates triggered single photon emission in the telecom C-band with low multi--photon probability of $g^{(2)}(0)=\num{0.11 \pm 0.03}$. The zero delay multi-photon probability $g^{(2)}(0)$--value is extracted by fitting the correlation peaks and comparing the area between the peak at zero time delay and the side peaks. Using our current excitation method, we can extract the emission time which sets the maximum duty cycle of our source to be $\SI{1.37}{\nano\second} \pm \SI{50}{\pico\second}$, where the error is assumed from the maximum timing uncertainty of our measurement setup. This value can be drastically reduced by performing resonant excitation of specific optical transitions\cite{Michler2009}, as any non-radiative contributions to the rise time are eliminated in this case. The rise and decay time are convoluted for non-resonant excitation, which can make the decay time appear artificially longer. Furthermore, the quantum dot lifetime can also be reduced by including the emitter into optical cavity structures, as demonstrated in \cite{Herzog2018, Kolatschek2019} for quantum dots grown by metal-organic vapor-phase epitaxy.

\begin{figure}[htbp]
\begin{center}
    \includegraphics[width=110mm]{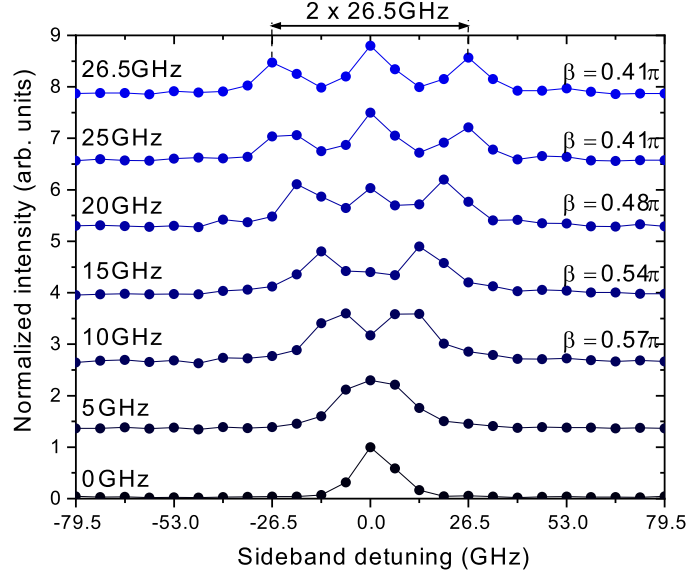}
    \caption{\label{fig:figure3} Tunable modulation of single photon carriers between 0 and \SI{26.5}{\giga\hertz} creating upper and lower sidebands with variable separation. The emission of the carrier line is at \SI{1547.21}{\nano\meter}. The data is recorded with a spectral resolution of \SI{35}{\pico\meter} or \SI{4.37}{\giga\hertz}. The modulation index is indicated on the right-hand side of the graph.} 
    \end{center}
\end{figure}

The ideal trion state for our experiment after the modulation would be a superposition state of three frequency channels $\ket{\mathrm{state}}=1/\sqrt{2}\ket{\mathrm{carrier}}+1/2\ket{\mathrm{carrier} + f}-1/2\ket{\mathrm{carrier} - f}$. We approximate this state using phase modulation with a single frequency that creates side-bands at multiples of $f$ following the ratio of Bessel functions of the first kind. Therefore we are interested into the first order sidebands and want to minimize the photons lost in higher order sidebands. Using single--frequency phase modulation of \SI{26.5}{\giga\hertz}, and a modulation index of $0.41 \pi$, we shift \SI{54.6}{\percent} of the photons into the higher (\SI{27.3}{\percent}) and lower (\SI{27.3}{\percent}) sidebands  ($\SI{1547.21}{\nano\meter} \pm \SI{26.5}{\giga\hertz}$) and loose \SI{7.1}{\percent} to higher order sidebands (Fig.~\ref{fig:figure1}(c)). The separation of the sidebands is tunable between 0 and \SI{26.5}{\giga\hertz} (Fig.~\ref{fig:figure3}) and can be aligned to a chosen channel spacing. This separation is mainly limited by the signal generator (\SI{26.5}{\giga\hertz}) and the modulator (\SI{40}{\giga\hertz}). Recent demonstrations by Mercante et. al. show modulations speeds up to \SI{500}{\giga\hertz} or \SI{\pm 4}{\nano\meter} at \SI{1550}{\nano\meter}\cite{Mercante2018}, demonstrating the potential of this standard technique applied to quantum states of light.

We include \SI{1.6}{\kilo\meter} of optical fiber between the phase modulator and the spectral filtering of the two sidebands and demonstrate our setup can bridge intra--data center distances.
Spectral filtering with the transmission spectrometer allows to separate the sidebands from each other, as shown in Fig.~\ref{fig:figure1}\,(d). The emission time setting the maximum duty cycle of the photons in lower and upper sideband are measured to be $\SI{1.39}{\nano\second} \pm \SI{50}{\pico\second}$ and $\SI{1.39}{\nano\second} \pm \SI{50}{\pico\second}$ respectively. This value is the same as for unmodulated photons within our measurement accuracy. To determine whether our modulation technique is altering the single photon emission of the quantum dot, we perform a cross--correlation measurement between the two sidebands. After spectrally separating the sidebands from each other we measure the time difference between the arrival of the photons in the sidebands. As they are generated from single photons, the peak at zero time delay should still vanish as in the initial measurement. The two sidebands show a correlation probability of $g^{(2)}(0)=\num{0.16 \pm 0.06}$, demonstrating single photons in both sidebands.

\section{Conclusion}
We showed the use of a triggered single--photon source based on MOVPE--grown InAs/GaAs quantum dots with standard telecom equipment in the C-band bridging \SI{1.6}{\kilo\meter} distance. The widely available equipment, already installed infrastructure and the low losses in fibers highlight the importance of this wavelength range for quantum communication. 
We created single photons with tunable sidebands (up to \SI{53}{\giga\hertz} separation) with a multi--photon emission probability in the sidebands of $g^{(2)}(0)=\num{0.16 \pm 0.06}$. This can be seen as a single photon source multiplexed on several channels or can be used as a resource for a frequency based qubit. With standard IQ modulators, a similar technique allows to shift the carrier\cite{Lo2017} to fine-tune the overlap with another source.

\newpage

\section*{Funding}
Swedish Research Council (VR): (875994, 2018-04812, 2016-03905, 2013-7152, 2016-04510); EU Horizon 2020: (820423, 749971); Knut and Alice Wallenberg Foundation; Dr. Isolde Dietrich Foundation;

\section*{Acknowledgments}
The authors acknowledge the contribution of Lucas Schweickert to the infrastructure of the optics lab. The authors thank the company Swabian instruments for providing their ``Time Tagger 20''. The authors thank Prof. Rinaldo Trotta for helpful discussions concerning the setup.

\bibliography{Sideband-creation-TK}

\begin{thebibliography}{10}
\newcommand{\enquote}[1]{``#1''}

\bibitem{DiVincenzo1995}
D.~P. DiVincenzo, \enquote{Quantum {{Computation}},}
  {\protect\JournalTitle{Science}} \textbf{270}, 255--261 (1995).

\bibitem{Kimble2008}
H.~J. Kimble, \enquote{The quantum internet,} {\protect\JournalTitle{Nature}}
  \textbf{453}, 1023--1030 (2008).

\bibitem{Bouwmeester1997}
D.~Bouwmeester, J.-W. Pan, K.~Mattle, M.~Eibl, H.~Weinfurter, and A.~Zeilinger,
  \enquote{Experimental quantum teleportation,} {\protect\JournalTitle{Nature}}
  \textbf{390}, 575--579 (1997).

\bibitem{Boschi1998}
D.~Boschi, S.~Branca, F.~De~Martini, L.~Hardy, and S.~Popescu,
  \enquote{Experimental {{Realization}} of {{Teleporting}} an {{Unknown Pure
  Quantum State}} via {{Dual Classical}} and {{Einstein}}-{{Podolsky}}-{{Rosen
  Channels}},} {\protect\JournalTitle{Physical Review Letters}} \textbf{80},
  1121--1125 (1998).

\bibitem{Pirandola2015}
S.~Pirandola, J.~Eisert, C.~Weedbrook, A.~Furusawa, and S.~L. Braunstein,
  \enquote{Advances in quantum teleportation,} {\protect\JournalTitle{Nature
  Photonics}} \textbf{9}, 641--652 (2015).

\bibitem{Reindl2018}
M.~Reindl, D.~Huber, C.~Schimpf, S.~F.~C. da~Silva, M.~B. Rota, H.~Huang,
  V.~Zwiller, K.~D. J\"ons, A.~Rastelli, and R.~Trotta, \enquote{All-photonic
  quantum teleportation using on-demand solid-state quantum emitters,}
  {\protect\JournalTitle{Science Advances}} \textbf{4}, eaau1255 (2018).

\bibitem{Sherson2006}
J.~F. Sherson, H.~Krauter, R.~K. Olsson, B.~Julsgaard, K.~Hammerer, I.~Cirac,
  and E.~S. Polzik, \enquote{Quantum teleportation between light and matter,}
  {\protect\JournalTitle{Nature}} \textbf{443}, 557--560 (2006).

\bibitem{Gao2013}
W.~B. Gao, P.~Fallahi, E.~Togan, A.~Delteil, Y.~S. Chin, J.~{Miguel-Sanchez},
  and A.~Imamo{\u g}lu, \enquote{Quantum teleportation from a propagating
  photon to a solid-state spin qubit,} {\protect\JournalTitle{Nature
  Communications}} \textbf{4}, 2744 (2013).

\bibitem{Miya1979}
T.~Miya, Y.~Terunuma, T.~Hosaka, and T.~Miyashita, \enquote{Ultimate low-loss
  single-mode fibre at 1.55 {$M$}m,} {\protect\JournalTitle{Electronics
  Letters}} \textbf{15}, 106--108 (1979).

\bibitem{Wehner2018}
S.~Wehner, D.~Elkouss, and R.~Hanson, \enquote{Quantum internet: {{A}} vision
  for the road ahead,} {\protect\JournalTitle{Science}} \textbf{362}, eaam9288
  (2018).

\bibitem{Agrawal2010}
G.~P. Agrawal, \emph{Fiber-Optic Communication Systems}, no. 222 in Wiley
  Series in Microwave and Optical Engineering ({Wiley}, 2010), 4th ed.

\bibitem{Chithrani2004}
D.~Chithrani, R.~L. Williams, J.~Lefebvre, P.~J. Poole, and G.~C. Aers,
  \enquote{Optical spectroscopy of single, site-selected, {{InAs}}/{{InP}}
  self-assembled quantum dots,} {\protect\JournalTitle{Applied Physics
  Letters}} \textbf{84}, 978--980 (2004).

\bibitem{Paul2017}
M.~Paul, F.~Olbrich, J.~H\"oschele, S.~Schreier, J.~Kettler, S.~L. Portalupi,
  M.~Jetter, and P.~Michler, \enquote{Single-photon emission at 1.55 {$M$}m
  from {{MOVPE}}-grown {{InAs}} quantum dots on {{InGaAs}}/{{GaAs}} metamorphic
  buffers,} {\protect\JournalTitle{Applied Physics Letters}} \textbf{111},
  033102 (2017).

\bibitem{Muller2018}
T.~M\"uller, J.~{Skiba-Szymanska}, A.~B. Krysa, J.~Huwer, M.~Felle,
  M.~Anderson, R.~M. Stevenson, J.~Heffernan, D.~A. Ritchie, and A.~J. Shields,
  \enquote{A quantum light-emitting diode for the standard telecom window
  around 1,550 nm,} {\protect\JournalTitle{Nature Communications}} \textbf{9},
  862 (2018).

\bibitem{He2013}
Y.-M. He, Y.~He, Y.-J. Wei, D.~Wu, M.~Atat\"ure, C.~Schneider, S.~H\"ofling,
  M.~Kamp, C.-Y. Lu, and J.-W. Pan, \enquote{On-demand semiconductor
  single-photon source with near-unity indistinguishability,}
  {\protect\JournalTitle{Nature Nanotechnology}} \textbf{8}, 213--217 (2013).

\bibitem{Akopian2006}
N.~Akopian, N.~H. Lindner, E.~Poem, Y.~Berlatzky, J.~Avron, D.~Gershoni, B.~D.
  Gerardot, and P.~M. Petroff, \enquote{Entangled {{Photon Pairs}} from
  {{Semiconductor Quantum Dots}},} {\protect\JournalTitle{Physical Review
  Letters}} \textbf{96}, 130501 (2006).

\bibitem{Olbrich2017}
F.~Olbrich, J.~H\"oschele, M.~M\"uller, J.~Kettler, S.~Luca~Portalupi, M.~Paul,
  M.~Jetter, and P.~Michler, \enquote{Polarization-entangled photons from an
  {{InGaAs}}-based quantum dot emitting in the telecom {{C}}-band,}
  {\protect\JournalTitle{Applied Physics Letters}} \textbf{111}, 133106 (2017).

\bibitem{Davanco2017}
M.~Davanco, J.~Liu, L.~Sapienza, C.-Z. Zhang, J.~V. D.~M. Cardoso, V.~Verma,
  R.~Mirin, S.~W. Nam, L.~Liu, and K.~Srinivasan, \enquote{Heterogeneous
  integration for on-chip quantum photonic circuits with single quantum dot
  devices,} {\protect\JournalTitle{Nature Communications}} \textbf{8}, 889
  (2017).

\bibitem{Zeuner2018}
K.~D. Zeuner, M.~Paul, T.~Lettner, C.~Reuterski\"old~Hedlund, L.~Schweickert,
  S.~Steinhauer, L.~Yang, J.~Zichi, M.~Hammar, K.~D. J\"ons, and V.~Zwiller,
  \enquote{A stable wavelength-tunable triggered source of single photons and
  cascaded photon pairs at the telecom {{C}}-band,}
  {\protect\JournalTitle{Applied Physics Letters}} \textbf{112}, 173102 (2018).

\bibitem{Senellart2017}
P.~Senellart, G.~Solomon, and A.~White, \enquote{High-performance semiconductor
  quantum-dot single-photon sources,} {\protect\JournalTitle{Nature
  Nanotechnology}} \textbf{12}, 1026--1039 (2017).

\bibitem{Lo2017}
H.-P. Lo and H.~Takesue, \enquote{Precise tuning of single-photon frequency
  using an optical single sideband modulator,} {\protect\JournalTitle{Optica}}
  \textbf{4}, 919--923 (2017).

\bibitem{Belthangady2010}
C.~Belthangady, C.-S. Chuu, I.~A. Yu, G.~Y. Yin, J.~M. Kahn, and S.~E. Harris,
  \enquote{Hiding {{Single Photons}} with {{Spread Spectrum Technology}},}
  {\protect\JournalTitle{Physical Review Letters}} \textbf{104}, 223601 (2010).

\bibitem{Kolchin2008}
P.~Kolchin, C.~Belthangady, S.~Du, G.~Y. Yin, and S.~E. Harris,
  \enquote{Electro-{{Optic Modulation}} of {{Single Photons}},}
  {\protect\JournalTitle{Physical Review Letters}} \textbf{101} (2008).

\bibitem{Specht2009}
H.~P. Specht, J.~Bochmann, M.~M\"ucke, B.~Weber, E.~Figueroa, D.~L. Moehring,
  and G.~Rempe, \enquote{Phase shaping of single-photon wave packets,}
  {\protect\JournalTitle{Nature Photonics}} \textbf{3}, 469--472 (2009).

\bibitem{Merolla2002}
J.-M. Merolla, L.~Duraffourg, J.-P. Goedgebuer, A.~Soujaeff, F.~Patois, and
  W.~Rhodes, \enquote{Integrated quantum key distribution system using single
  sideband detection,} {\protect\JournalTitle{The European Physical Journal D -
  Atomic, Molecular and Optical Physics}} \textbf{18}, 141--146 (2002).

\bibitem{Lukens2017}
J.~M. Lukens and P.~Lougovski, \enquote{Frequency-encoded photonic qubits for
  scalable quantum information processing,} {\protect\JournalTitle{Optica}}
  \textbf{4}, 8--16 (2017).

\bibitem{Kues2017}
M.~Kues, C.~Reimer, P.~Roztocki, L.~R. Cort\'es, S.~Sciara, B.~Wetzel,
  Y.~Zhang, A.~Cino, S.~T. Chu, B.~E. Little, D.~J. Moss, L.~Caspani,
  J.~Aza\~na, and R.~Morandotti, \enquote{On-chip generation of
  high-dimensional entangled quantum states and their coherent control,}
  {\protect\JournalTitle{Nature}} \textbf{546}, 622--626 (2017).

\bibitem{Paudel2018}
U.~Paudel, A.~P. Burgers, D.~G. Steel, M.~K. Yakes, A.~S. Bracker, and
  D.~Gammon, \enquote{Generation of frequency sidebands on single photons with
  indistinguishability from quantum dots,} {\protect\JournalTitle{Physical
  Review A}} \textbf{98}, 011802 (2018).

\bibitem{Schweickert2018}
L.~Schweickert, K.~D. J\"ons, K.~D. Zeuner, S.~F. {Covre da Silva}, H.~Huang,
  T.~Lettner, M.~Reindl, J.~Zichi, R.~Trotta, A.~Rastelli, and V.~Zwiller,
  \enquote{On-demand generation of background-free single photons from a
  solid-state source,} {\protect\JournalTitle{Applied Physics Letters}}
  \textbf{112}, 093106 (2018).

\bibitem{Michler2009}
P.~Michler, ed., \emph{Single Semiconductor Quantum Dots}, Nanoscience and
  Technology ({Springer}, 2009). OCLC: ocn248993461.

\bibitem{Herzog2018}
T.~Herzog, M.~Sartison, S.~Kolatschek, S.~Hepp, A.~Bommer, C.~Pauly,
  F.~M\"ucklich, C.~Becher, M.~Jetter, S.~L. Portalupi, and P.~Michler,
  \enquote{Pure single-photon emission from {{In}}({{Ga}}){{As QDs}} in a
  tunable fiber-based external mirror microcavity,}
  {\protect\JournalTitle{Quantum Science and Technology}} \textbf{3}, 034009
  (2018).

\bibitem{Kolatschek2019}
S.~Kolatschek, S.~Hepp, M.~Sartison, M.~Jetter, P.~Michler, and S.~L.
  Portalupi, \enquote{Deterministic fabrication of circular {{Bragg}} gratings
  coupled to single quantum emitters via the combination of in-situ optical
  lithography and electron-beam lithography,} {\protect\JournalTitle{Journal of
  Applied Physics}} \textbf{125}, 045701 (2019).

\bibitem{Mercante2018}
A.~J. Mercante, S.~Shi, P.~Yao, L.~Xie, R.~M. Weikle, and D.~W. Prather,
  \enquote{Thin film lithium niobate electro-optic modulator with terahertz
  operating bandwidth,} {\protect\JournalTitle{Optics Express}} \textbf{26},
  14810--14816 (2018).

\end{thebibliography}

\end{document}